\documentclass[aps,pra,showpacs,twocolumn,superscriptaddress]{revtex4}
\usepackage{graphicx}
\usepackage{bm}
\usepackage{mathrsfs}

\begin{document}
\title{Spin Excitation Spectra of Spin-Orbit Coupled Bosons in Optical Lattice}
\author{R.-Y. Li}
\affiliation{School of Physics, Peking University,
Beijing 100048, China}
\author{Liang He}
\affiliation{Institut f\"{u}r Theoretische Physik, Leopold-Franzens
Universit\"{a}t Innsbruck, A-6020 Innsbruck, Austria}
\author{Q. Sun}
\affiliation{Department of Physics, Capital
Normal University, Beijing 100048, China}
\author{A.-C. Ji}
\affiliation{Department of Physics, Capital
Normal University, Beijing 100048, China}
\author{G.-S. Tian}
\affiliation{School of Physics, Peking University,
Beijing 100048, China}

\date{{\small \today}}

\begin{abstract}
Spin-wave excitation plays important roles in the investigation of
the magnetic phases. In this paper, we study the spin-wave
excitation spectra of two-component Bose gases with spin-orbit
coupling on a deep square optical lattice using spin-wave theory. We
find that, while the excitation spectrum of the vortex crystal phase
is gapless with a linear dispersion in the vicinity of the minimum
point, the spectra of the commensurate spiral spin phase and the
skyrmion crystal phase are gapped. Significantly, the spin
fluctuations strongly destabilize the classical ground state of the
skyrmion phase. It suggests the emergence of a new state in the
phase diagram. Such features of the spin excitation spectra provide
further insights into the exploration of the exotic spin phases.
\end{abstract}









\maketitle

\section {INTRODUCTION}

Typical spin-orbit coupling (SOC) of electrons in solid state system
is due to relativistic effect. For instance, it can be derived from
the Dirac equation of fermions by expansion in terms of the small
parameter $v/c\approx 0.01$, which is the ratio of electron and
light speeds in solids. Therefore, the spin-orbit interaction is
much weaker than the Coulomb interaction and is generally neglected
in the study of condensed matter physics. However, it has been
recently noticed that the SOC plays a crucial role in leading to the
so-called topological insulating states in solids \cite{Hasan,Qi}.
Consequently, it attracts many physicists' attention. In particular,
with successful realization of the synthetic gauge field in
ultracold quantum gases \cite{Lin1,Lin2,Zhang1,Wang1,Cheuk}, it has
been discovered that the spin-orbit coupling may be also important
in determining the properties of these systems
\cite{Wang2,Ho,XuZF1,CongjunWu,Zhang,Li,Sinha,Hu1,Wilson,Kawakami,
Su,Deng,Stanescu,Vyasanakere,Hu2,Yu,Gong}. For example, SOC can
produce various new types of ground states such as the stripes, the
half-vortex phases in the Bose-Einstein condensates, and a bounded
state, which is called Rashbon in the Fermi gases
\cite{XFZhou,NGoldman,HZhai}. That makes research on SOC in
ultracold gases vigorous.

In addition, ultracold gases on optical lattices have attracted
great interests to simulate  a wide range of  condensed-matter
phenomena \cite{Lewenstein}, such as the  the Mott metal-insulator
transition and the effects of strong correlation in the Hubbard
model \cite{Esslinger}.  As is known, at half filling, the low
energy physics of these systems can be indeed described by an
effective antiferromagnetic spin Hamiltonian in the deep Mott regime
\cite{Anderson,Datta,Wang3,Ji}. Therefore, the ground states are
generally Neel ordered and the corresponding low-lying excitations
are antiferromagnetic magnons.

Then, a natural question which one would like to ask is what effects
will be induced if SOC exists in ultracold gases loaded in optical
lattice. Actually, in this case, one can derive an additional
Dzyaloshinskii-Moriya (DM) type of super-exchange interaction in the
deep Mott insulating regime by the second-order perturbation theory
\cite{Dzyaloshinsky,Moriya}. Very recently, many authors analyzed
the phase diagram of this  system \cite{Cole,Radic,Cai,MGong,YQian}.
By applying the classical Monte-Carlo simulations, several exotic
spin-textures, whose existences were previously explored in various
solid state materials with the DM interaction
\cite{Pesin,Heinze,Banerjee}, are also found in ultracold gases.

However, these investigations are not complete from the theoretical
point of view. Indeed, up to now, only the classical ground-state
phase diagram of the bosonic super-exchange Hamiltonian on a square
lattice was numerically explored. But, the effects of spin
fluctuations on the stabilities of exotic spin-textures in these
phases have been not discussed yet. As is well known, strong spin
fluctuations in low-dimensional systems can qualitatively change
their phase diagrams outlined by the classical simulations.
Therefore, one has to take these fluctuations into consideration in
order to determine the stability of each possible phase. In the
following, we shall proceed to calculate the low-lying spectra of
spin fluctuations in these phases. As a result, we are able to
determine the parameter region in which each corresponding phase is
stable against spin fluctuations.

Technically, we implement the spin-wave theory developed in
Ref.~\cite{Kleine} to derive the global magnetic phase diagram of
the super-exchange Hamiltonian. In particular, we calculate the
spin-wave excitation spectra for the $XY$-ferromagnet phase, the
spiral spin phase and the  spin vortex phase, whose existences were
studied by the previous classical Monte Carlo simulations
\cite{Pesin,Heinze,Banerjee}. We discuss also the stability of a
novel phase, the so-called skyrmion crystal phase in bosonic
ultracold gases. We show that the possible existing regions for the
skyrmion crystal phase are greatly shrunk by the spin fluctuations.
More precisely, we find that the spin excitation spectrum of the
commensurate spiral spin phase has two minima, which are symmetrical
with respect to the $k_y$-axis in momentum space. In addition, it
possesses a gap, too. On the other hand, the spectrum of the
$3\times3$ skyrmion crystal phase presents some intriguing features,
which render its existence fragile. These characteristic excitation
spectra may be investigated in the future experiment. Given that the
above exotic spin configurations have also been discovered and
attracted great interests in recent solid state materials
\cite{Pesin,Heinze,Banerjee},such features of the spin excitation
spectra, which should be observable via the neutron scattering
experiments, provide insights into these spin phases.

This paper is organized as follows. In section II, we derive the
effective super-exchange spin model for the systems of bosonic gases
on a square optical lattice; In section III, we present the general
formalism of the spin wave theory developed in Ref.~\cite{Kleine};
Then, in section IV, we analyze the phase diagram and excitation
spectrum of each above-mentioned phase in detail; Finally, in
section V, we summarize our main results and discuss some related
issues with experiments.

\section {THE EFFECTIVE SPIN MODEL}

Let us consider a boson gas loaded in a two dimensional square
optical lattice. The atoms are characterized by two
nearly-degenerate states, which are produced by a hyper-fine
interaction. In the following, for convenience, we shall use the
language of (pseudo)-spin to describe the local occupancy of these
states by particles. Furthermore, we take the synthetic spin-orbit
interaction into consideration. Then, the Hamiltonian of the bosonic
gas can be written as
\begin{equation}
H=-t\sum_{\langle i,j\rangle}(\psi^\dag_i\mathcal{R}_{ij}\psi_j+{\rm h.c.})
+\frac{1}{2}\sum_{i\sigma\sigma'}U_{\sigma\sigma'}
a^\dag_{i\sigma}a^\dag_{i\sigma'}a_{i\sigma'}a_{i\sigma}.
\end{equation}
Here $a_{i\sigma}$ ($a^\dag_{i\sigma}$) denotes boson annihilation
(creation) operator, which annihilates (creates) a boson of
(pseudo)-spin $\sigma=\uparrow,\downarrow$ at lattice site $i$. In a
compact form, these operators can be re-written as
$\psi^\dag_i\equiv(a^\dag_{i\uparrow},a^\dag_{i\downarrow})$ and its
Hermitian  conjugate $\psi_i$. The first term of $H$ describes the
hopping process of bosons between a pair of nearest-neighbor lattice
sites $i$ and $j$. The corresponding hopping matrices are given by
$\mathcal{R}_{ij}\equiv\exp[i\vec{A}\cdot(\vec{r}_i-\vec{r}_j)]$,
where $\vec{A}=(\alpha\sigma_y,\beta\sigma_x,0)$ is a non-Abelian
gauge field  with $\sigma_x$ and $\sigma_y$ being the Pauli matrices
\cite{Cole,Radic,Goldman}. In fact, if we let $\beta=-\alpha$, then
$\vec{A}=(\alpha\sigma_y,-\alpha\sigma_x,0)$ is the gauge field
giving rise to the well-known Rashba SOC.  We see that the diagonal
part of the hopping matrices $\mathcal{R}_{ij}$ represents the
spin-conserved tunneling, while the off-diagonal part is for the
spin-flipped tunneling of bosons between the lattice sites $i$ and
$j$. Furthermore, the second term in the Hamiltonian stands for the
contact interaction between bosons. In general, the intra-species
interaction strengths can be different. However we assume them to be
equal in the following discussions. More precisely, we set
$U_{\uparrow\uparrow}=U_{\downarrow\downarrow}=U$. Similarly,  we
choose the inter-species interaction
$U_{\uparrow\downarrow}=U_{\downarrow\uparrow}\equiv\lambda{U}$ with
$\lambda>0$ being a dimensionless parameter.

As is well known, the filling factor of bosons is also an important
parameter to determine the properties of ultracold gases
on an optical lattice. In the following, we shall only consider
the case of one boson per site on average. Now, we can perform
the standard Schrieffer-Wolf transformation
$H_{\rm eff}=e^{iS}He^{-iS}$ \cite{Hewson} to the Hamiltonian
in the large-$U$ limit $U_{\sigma\sigma^\prime}\gg{t}$.
It produces an effective Hamiltonian
\begin{equation}
H_{\rm eff}=\sum_{i,\delta=\hat{x},\hat{y}}
\left[\sum_{a=x,y,z}J^a_\delta{}S^a_iS^a_{i+\delta}+\bm{D}_\delta
\cdot(\bm{S}_i\times\bm{S}_{i+\delta})\right],
\label{effective Hamiltonian}
\end{equation}
which captures the low-energy physics of the original system
\cite{Duan,Cole,Radic}. Here, ${\bf S}_{i}=(S^x_i,S^y_i,S^z_i)$ are
the spin operators of bosons at lattice site $i$. In terms of the
boson operators $a_{i\sigma}$ and $a_{i\sigma}^\dagger$, they can be
written as $S^x_i=(a_{i\uparrow}^\dagger
a_{i\downarrow}+a_{i\downarrow}^\dagger a_{i\uparrow})/2$,
$S^y_i=(a_{i\uparrow}^\dagger
a_{i\downarrow}-a_{i\downarrow}^\dagger a_{i\uparrow})/2i$ and
$S^z_i=(n_{i\uparrow} - n_{i\downarrow})/2$. In Eq.~(\ref{effective
Hamiltonian}), the first summation is a Heisenberg type of
Hamiltonian and the second summation represents the so-called
Dzyaloshinskii-Moriya (DM) superexchange interaction. The
corresponding coupling constants are given in TABLE I.
\begin{table}[h]
\centering
\begin{tabular}{ll}
\hline\hline
  $J^x_{\hat{x}}=-\frac{4t^2}{\lambda{}V}\cos(2\alpha)$& $J^x_{\hat{y}}
  =-\frac{4t^2}{\lambda{}V}$\\ \hline
  $J^y_{\hat{x}}=-\frac{4t^2}{\lambda{}V}$  &  $J^y_{\hat{y}}
  =-\frac{4t^2}{\lambda{}V}\cos(2\alpha)$\\ \hline
  $J^z_{\hat{x}}=-\frac{4t^2}{\lambda{}V}(2\lambda-1)\cos(2\alpha)$ &
  $J^z_{\hat{y}}=-\frac{4t^2}{\lambda{}V}(2\lambda-1)\cos(2\alpha)$\\ \hline
  ${\bf D}_{\hat{x}}=\frac{4t^2}{V}\sin(2\alpha){\bf e}_y$ & ${\bf D}_{\hat{y}}
  =-\frac{4t^2}{V}\sin(2\alpha){\bf e}_x$\\ \hline\hline
\end{tabular}
\caption{Exchange couplings $J^a_\delta$ for the Heisenberg term
and $\bm{D}_\delta$ for the DM superexchange interaction
in the effective Hamiltonian.}
\label{table}
\end{table}

We would like to emphasize that these interactions favor
qualitatively different types of spin orders. In fact, the
Heisenberg Hamiltonian gives either the ferromagnetic or the
anti-ferromagnetic ordering, while the DM interaction makes the
spiral spin ordering possible. In other words, there exists a
competition between them. Consequently, their interplay may drive
the system to various exotic phases. In the following, we shall
first determine the possible existence of several phases in the
system and then, analyze their stabilities by applying the spin-wave
theory.
\begin{figure}[h]
\includegraphics[width=0.48\textwidth]{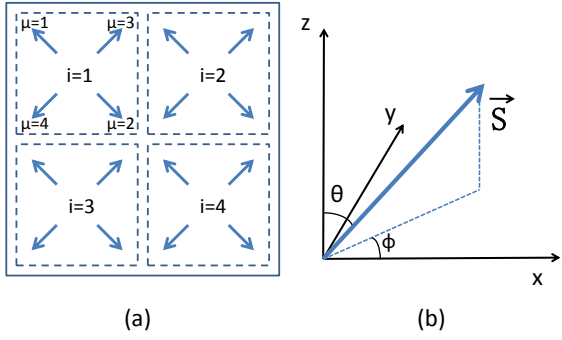}
\caption{(a) Spin configuration of the $2\times2$ spin vortex phase,
where $i$ denotes the magnetic primitive cell and $\mu$ is the
sublattice index. (b) The spin is rotated from the $z$ axis  to a
new direction ($\theta,\phi$) in the spin-wave theory. \label{fig1}}
\end{figure}

\section{FORMULISM OF THE SPIN WAVE THEORY}

In the previous works, some authors have explored the classical
ground states of Hamiltonian (\ref{effective Hamiltonian}) by the
classical Monte Carlo simulations \cite{Cole,Radic}. They showed
that the classical ground states of this system support a variety of
phases such as the $XY$-ferromagnetic ($XY$-FM) phase, the
longitudinal ferromagnetic ($Z$-FM) and antiferromagnetic ($Z$-AFM)
phases, the spiral spin phase, the $3\times3$ skyrmion crystal (SkX)
phase, and the $2\times 2$ spin vortex (VX) phase as the coupling
constants of Hamiltonian (\ref{effective Hamiltonian}) are changed.
However, they did not consider the stabilities of these phases. In
order to address this issue, one has to take the strong spin
fluctuations of the system into consideration, as we do in the
following.

To start with, we shall first calculate the  ground state energy of
each above-mentioned phase in some specified region of parameters by
the mean-field theory and see if the previous phase diagram can be
re-established. Then, we compute the second-order correction to the
energies of those phases. This correction is due to  the spin-wave
excitations. In some cases, it can render the phase derived by the
classical methods invalid. More precisely, as we shall show in the
following, the excitation spectrum of a specified phase may develop
an imaginary part and hence, becomes a complex function of momentum
in some parameter regions.  It implies that the  assumed phase is
actually unstable against the spin fluctuations in these regions. To
illustrate the procedure, let us take the $2\times2$ VX phase for
example.

In the first step, we perform the following transformation
\begin{equation}
[S^x_{\mu,i},S^y_{\mu,i},S^z_{\mu,i}]^T=\hat{A}_\mu[\widetilde{S}^x_{\mu,i},
\widetilde{S}^y_{\mu,i},\widetilde{S}^z_{\mu,i}]^T,
\end{equation}
where $i$ denotes the spin primitive cell of the specified phase and
$\mu$ is the sublattice index, as shown in Fig.~(\ref{fig1}a), to
$H_{\rm eff}$. The matrix $\hat{A}_\mu$  reads
\begin{widetext}
\begin{equation}
\hat{A}_\mu=\left(\begin{array}{ccc}\cos\theta_\mu\cos^2\phi_\mu+\sin^2\phi_\mu
& \cos\phi_\mu\sin\phi_\mu(\cos\theta_\mu-1) &
\sin\theta_\mu\cos\phi_\mu \\
\cos\phi_\mu\sin\phi_\mu(\cos\theta_\mu-1) &
\cos\theta_\mu\sin^2\phi_\mu+\cos^2\phi_\mu & \sin\theta_\mu\sin\phi_\mu\\
-\sin\theta_\mu\cos\phi_\mu & -\sin\theta_\mu\sin\phi_\mu & \cos\theta_\mu \\
\end{array}\right).
\end{equation}
\end{widetext}
It represents the rotation of each spin in the Hamiltonian from the
local $z$-axis to a new direction ($\theta,\phi$), as shown in
Fig.~(\ref{fig1}b). Next, we apply Holstein-Primakoff transformation
\cite{Holstein} by substituting the spin operators
$\widetilde{S}^x_{\mu,i}=(b_{\mu,i}+b^\dag_{\mu,i})/2,
\widetilde{S}^y_{\mu,i}=(b_{\mu,i}-b^\dag_{\mu,i})/(2i),
\widetilde{S}^z_{\mu,i}=1/2-b^\dag_{\mu,i}b_{\mu,i}$ into the
effective Hamiltonian. Consequently, we obtain $H_{\rm eff}\simeq
H_0+H_2$. Here, $H_0$ denotes the mean-field energy for a specific
phase and $H_2$ is the spin-wave Hamiltonian which describes the
spin fluctuation in the system. Let $N$ be the total number of
lattice sites, the mean-field energy $H_0$ can be written as
\begin{widetext}
\begin{eqnarray}
H_0 & = &
\frac{N}{16}\sum_{(\mu,\nu)}\left[\cos\phi_{\nu}\sin\theta_\nu
(-|\bm{D}_{\hat{x}}|\cos\theta_\mu +\cos\phi_\mu\sin\theta_\mu
J^x_{\hat{x}}) +
\sin\theta_\mu\sin\theta_\nu\sin\phi_\mu\sin\phi_\nu
J^y_{\hat{x}} \right. \nonumber \\
& + & \left.
\cos\theta_\nu(|\bm{D}_{\hat{x}}|\cos\phi_\mu\sin\theta_\mu
+\cos\theta_\mu J^z_{\hat{x}}) \right] +
\frac{N}{16}\sum_{(\alpha,\beta)}
\left[\sin\theta_\alpha\sin\phi_\alpha(-|\bm{D}_{\hat{y}}|
\cos\theta_\beta \right.\nonumber \\
& + & \left. \sin\theta_\beta\sin\phi_\beta J^x_{\hat{y}}) +
\cos\phi_\beta \cos\psi_\alpha\sin\theta_\beta\sin\theta_\alpha
J^y_{\hat{y}} +\cos\theta_\alpha(|\bm{D}_{\hat{y}}|\sin\theta_\beta
\sin\phi_\beta+\cos\theta_\beta J^z_{\hat{y}}) \right]. \label{H0}
\end{eqnarray}
\end{widetext}
Here, the spin angles $\theta_\mu$ and $\phi_\mu$  are determined by
the spin configurations of the specified mean-field ground state.
For example, in the $2\times2$ VX phase, we have
$\theta_1=\theta_2=\theta_3=\theta_4=\pi/2,\phi_1=3\pi/4,\phi_2=7\pi/4,
\phi_3=\pi/4$ and $\phi_4=5\pi/4$. In addition, $(\mu,\nu)$ and
$(\alpha,\beta)$ denote two pairs of nearest-neighbor sites in the
primitive cell along the $\hat{x}$- and $\hat{y}$-axis,
respectively. With the same notations, we obtain the following
spin-wave interaction terms
\begin{widetext}
\begin{equation}
H_2=\sum_\mu{}M_{\mu\mu}\sum_in_{\mu,i}+\sum_{\mu,\mu'}\sum_{<ij>}
(M_{\mu,\mu'}b^\dag_{\mu,i}b_{\mu',j}+N_{\mu,\mu'}
b^\dag_{\mu,i}b^\dag_{{\mu',j}}+{\rm h.c.}),
\label{H2}
\end{equation}
\end{widetext}
by keeping only the terms of two-boson operators in the expansion of
the transformed effective Hamiltonian. $M_{\mu,\mu'}$ and
$N_{\mu,\mu'}$ are the coefficient matrices depending on the
mean-field spin configuration. The explicit expressions of the
matrices are tedious and not given here. Then, by using the Fourier
transformation, we are able to re-write $H_2$ as
\begin{equation}
H_2=\frac{1}{2}\sum_{\vec{k}}\beta_{\vec{k}}
H_{MN}(\vec{k})\beta^\dag_{\vec{k}}+{\rm const},
\end{equation}
where $\beta_{\vec{k}}=(b^\dag_{\vec{k},1},b^\dag_{\vec{k},2},
b^\dag_{\vec{k},3},b^\dag_{\vec{k},4},b_{-\vec{k},1},
b_{-\vec{k},2},b_{-\vec{k},3},b_{-\vec{k},4})$ are boson operators
in the momentum space. Now, we apply the Bogoliubov transformation
$\beta_{\vec{k}}=\hat{T}\alpha_{\vec{k}}$ to the effective
Hamiltonian $H_2$. Due to the bosonic commutation relation, the
transformation matrix $\hat{T}$ must be canonical, i.e., relation
$\hat{T}^\dag\Lambda\hat{T}=\Lambda$, where
\begin{equation}
\Lambda=
\left(
\begin{array}{cc}
1_{4\times4}&0\\
0&-1_{4\times4}\\
\end{array}
\right).
\end{equation}
should hold true. Furthermore, after substituting the Bogoliubov
transformation into Eq.~(\ref{H2}), we determine the transformation
matrix $\hat{T}$ by requiring that the Hamiltonian be diagonalized
by it. In other words, the matrix product
$\hat{T}^\dag H_{MN}\hat{T}\equiv\hat{\omega}$
should be a diagonal matrix. With the canonical constraint
$\hat{T}^\dag\Lambda\hat{T}=\Lambda$, this condition
can be put into an equivalent form
\begin{equation}
(\Lambda\hat{T}\Lambda)^{-1}H_{MN}\Lambda(\Lambda\hat{T}\Lambda)
=\hat{\omega}\Lambda.
\end{equation}
Consequently, by diagonalizing $H_{MN}\Lambda$ with an invertible matrix
$\hat{V}=\Lambda\hat{T}\Lambda$, we can derive the spin excitation spectrum
$\omega_{\vec{k},\mu}$ and then, discuss the stability of the $2\times 2$
VX phase by seeing whether the calculated excitation spectral function
is complex.

In the following section, we shall study the possible existences and
stabilities of the $XY$-FM phase, $Z$-FM phase, $Z$-AFM phases, the
spiral spin phase, $3\times3$ SkX phase, and $2\times 2$ VX phase in
ultracold gases on a square optical lattice by the above-outlined
procedure. As shown below, although this procedure can be easily
applied to most of these phases, it is rather involved for the
$3\times3$ skyrmion crystal phase. The main difficulty is due to the
complicated spin configuration of this phase. It contains nine
sublattices. In other words, according to the local spin
orientations in the $3\times 3$ skyrmion phase, the whole lattice
can be divided into nine separate sublattices. In each of them, all
the localized spins point in a specified direction. A primitive spin
cell of this phase is shown in Fig.~\ref{fig2} (b).

\section{ANALYSIS OF THE SPIN EXCITATION SPECTRA}

First, we compute the ground-state energy $H_0$ for each possible
spin configuration listed in the last section and determine the phase diagram
by choosing the configuration of the lowest energy in different parameter regions.
Then, we shall compare our results with the previous ones derived
by the classical Monte Carlo calculations. Our phase diagram is shown
by the solid lines in Fig.~\ref{fig2} (a). It is in good agreement
with the Monte-Carlo simulations given in Ref.~\cite{Cole}.

Next, we calculate the spin-wave excitation spectrum of each phase
and see if it is stable against the spin fluctuations. In fact, we
find that the $Z$-AFM and $2\times2$ VX phases are stable. However,
the $3\times3$ SkX phase is pronouncedly affected by the spin
fluctuations. Actually, in two parameter regions, which we paint as
white areas in Fig.~\ref{fig2} (a), the spin-wave excitation
spectral functions have imaginary parts and hence, render the
corresponding phases unstable. On the left side of commensurate
4$\times$1 spiral phase, the unstable parameter region implies the
emergence of other commensurate spiral orders or incommensurate
spiral phases. While the right unstable parameter region may support
new state, which largely suppresses the $3\times3$ SkX phase.

Now, let us analyze the spin excitation spectrum
of each phase in a more detailed manner.

\subsection{$XY$-FM and $Z$-FM Phases}

We first consider the weak spin-orbit coupling regime in which the
parameter $\alpha$ in TABLE I is small. In this case, the Heisenberg
interaction in Hamiltonian (\ref{effective Hamiltonian}) is dominant
and the DM interaction can be treated as a perturbation. In
particular, if we set $\alpha=0$, then the effective Hamiltonian
becomes the standard ferromagnetic $XXZ$ model on a two-dimensional
square lattice \cite{Sachdev}. The phase diagram of this model is
well known: For $\lambda<1$, all the spins are aligned in the $XY$
plane. More precisely, the system is in the $XY$-FM phase. To this
phase, the Holstein-Primakoff transformation can be easily applied.
\begin{figure}[h]
\includegraphics[width=0.48\textwidth]{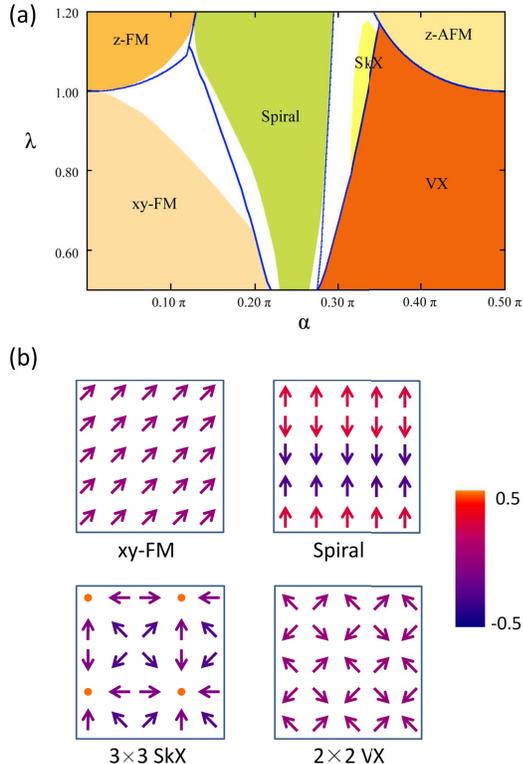}
\caption{(a) Phase diagram in the Mott insulating regime. The solid
lines are the mean-filed results. The filled areas with color are
the stable regions,  while the white areas are unstable ones under
the spin-wave fluctuations. (b) Spin configurations of  the $XY$-FM,
commensurate 4$\times$1 spiral spin, $3\times3$ SkX, and $2\times2$
VX phases. \label{fig2}}
\end{figure}
A direct calculation yields
$E(\vec{k})=\sqrt{\frac{1-\lambda}{2}}Ja\sqrt{k_x^2+k_y^2}$ for its
spin excitation spectrum, in which $J=\frac{4t^2}{\lambda{}V}$ and
$a$ is the lattice constant. Obviously, the excitation spectrum is
gapless with a linear dispersion near $\vert\vec{k}\vert\sim 0$,
which corresponds to the Goldstone modes in the $XY$-ferromagnets.
On the other hand, for $\lambda>1$, the system is in the $Z$-FM
phase. The spin excitation spectrum of this phase is given by
$E(\vec{k})=\Delta+\frac{Ja^2}{4}(k_x^2+k_y^2)$, which has a spin
gap at $\Delta=J(2\lambda-2)$. Finally, $\lambda=1$ is the so-called
Heisenberg point at which the model becomes the isotropic Heisenberg
ferromagnet. It has a gapless quadratic excitation spectrum, which
is shown in Fig.~\ref{fig3}.

\begin{figure}[h]
\includegraphics[width=0.4\textwidth]{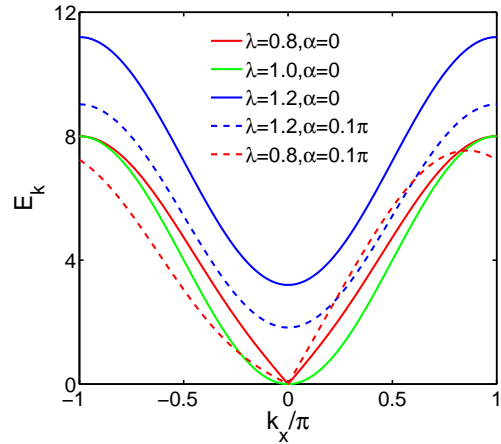}
\caption{The excitation spectra for the $XY$-FM ($\lambda=0.8$),
$Z$-FM ($\lambda=1.5$), and the Heisenberg ferromagnetic
($\lambda=1$) phases. Solid and dashed lines are for the strength
of SOC $\alpha=0$ and $\alpha=0.1\pi$, respectively.
\label{fig3}}
\end{figure}

\begin{figure*}[tbp]
\includegraphics[width=2\columnwidth\vspace{0cm}
\hspace{0cm}]{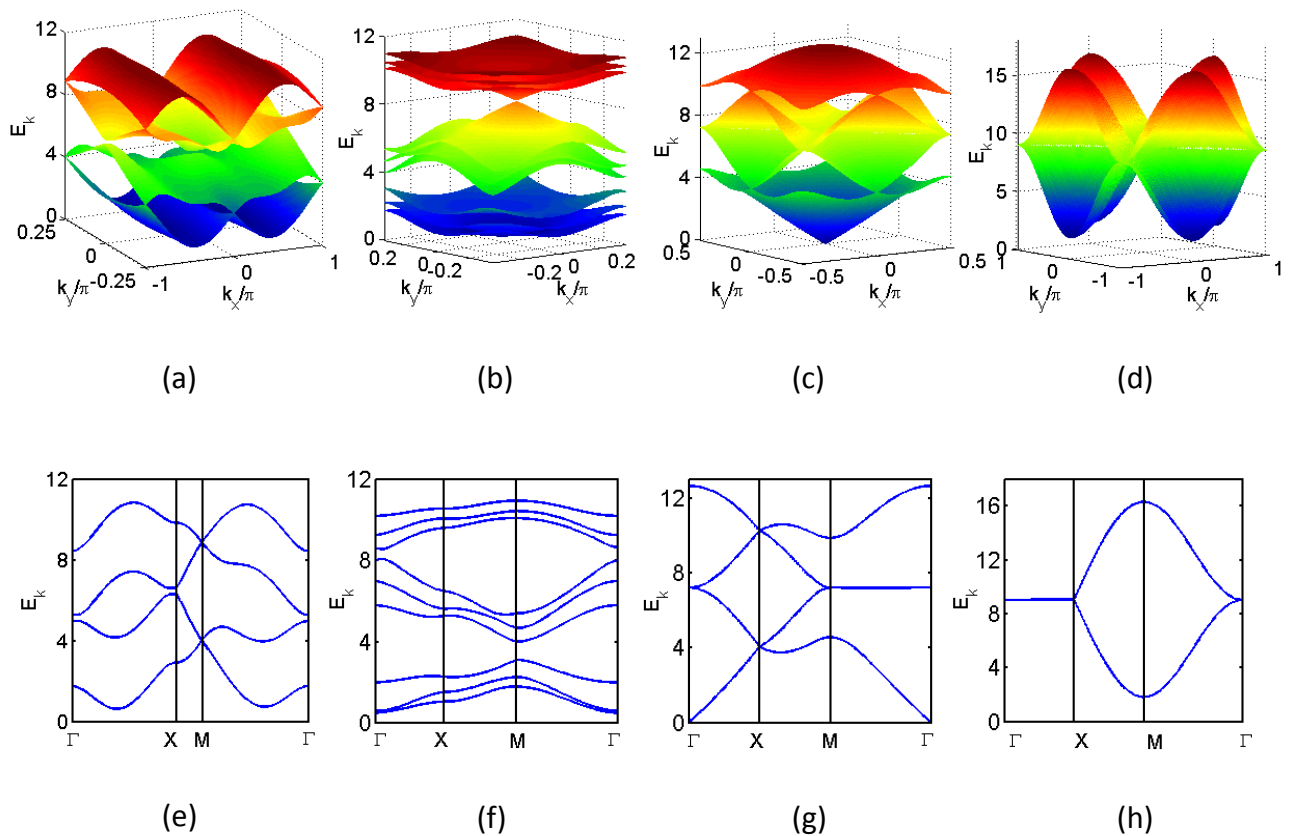} \caption{(Color online) The excitation
spectra for (a,e) 4$\times$1 spiral spin phase
($\lambda=0.8,\alpha=0.24\pi$). (b,f) 3$\times$3 skyrmion crystal
phase ($\lambda=0.9,\alpha=0.32\pi$), (c,g) 2$\times$2 spin vortex
phase ($\lambda=0.8,\alpha=0.4\pi$), and (d,h) Z-AFM
($\lambda=1.2,\alpha=0.4\pi$) phases. \label{fig4}}
\end{figure*}

Next, we take the effect of the DM interaction into consideration.
We find that the excitation spectrum remains still gapless and
linear near $\vert\vec{k}\vert\sim 0$ for the $XY$-FM phase. But, it
becomes asymmetric with respect to $k_x =0$ as shown by the dashed
line of $\lambda=0.8$ and $\alpha=0.1\pi$ in Fig.~\ref{fig3}.
Similar dispersion occurs along the  $k_y$ axis. In the meantime,
the gap of the $Z$-FM phase is decreased upon the increasing
strength of SOC. However, both the spin excitation spectra of the
$XY$-FM and $Z$-FM phases will become imaginary as the SOC strength
$\alpha$ is further enhanced. It indicates that these phases are
destabilized. This region of parameters is characterized by the
white area  on the left side of spiral phase in Fig.~\ref{fig2} (a).

\subsection{Spiral Spin Phase}

In the middle region of Fig.~\ref{fig2} (a), the DM interaction is
strong, comparing with the Heisenberg interaction. The competition
between these interactions leads to a spiral spin phase. To
determine the stability of this phase, we consider a commensurate
4$\times$1 periodic structure with all the spins lying in the $z-y$
plane. Notice that, in this configuration, the angles $\theta_i$
between the spins and $z$ axis are specified by $\theta_1=\theta_2$,
$\theta_3=\theta_4$, and $\theta_1+\theta_3=\pi$.

Then, we calculate the excitation spectrum of this phase by the
spin-wave analysis. Our results are shown in Fig.~\ref{fig4} (a) and
Fig.~\ref{fig4} (e). One can easily see that there exists a spin gap
in the diagram. In other words, the Goldstone mode is absent in this
phase. Moreover, the gap has two symmetric minima at
($\pm{}k_{0},0$), as shown in Fig.~\ref{fig4} (a). It is due to the
fact that the spin configuration in the spiral spin phase is
symmetric with respect to the $y$-axis. By varying parameter
$\alpha$, we find that  the energy gap gradually closes as the DM
interaction increases and the excitation spectrum becomes linear in
the vicinity of these minimal points. At the same time, the
4$\times$1 periodic structure becomes unstable. It suggests the
emergence of of other commensurate spiral orders or incommensurate
spiral phases.

\subsection{$2\times2$ VX and $Z$-AFM Phases}

Next, let us consider the large spin-orbit coupling regime around
$\alpha=\pi/2$, in which the spin-conserving hopping of particles is
suppressed. The Wilson loop, which is characterized by $W={\rm
tr}[R_{x}R_{y}R_{x}^\dagger R_{y}^\dagger]$ is reduced to a
marginally abelian one with $|W|=2$ \cite{Goldman}. In this
situation, the Heisenberg interaction is dominant and the spin
configuration is determined by a uniform antiferromagnetic
interaction in the $Z$-direction plus anisotropic spin-interactions
in $X$- and $Y$-directions. When parameter $\lambda<1$, the spin
interaction in the $XY$-plane is stronger. Therefore, the system has
a co-planar magnetic order. On the other hand, since the spin
interaction is antiferromagnetic in the $X$-direction but
ferromagnetic in the $Y$-direction, the classical ground state of
the system becomes a $2\times 2$ vortex crystal. Further analysis
yields the spin-wave spectrum of this phase, as shown in
Fig.~\ref{fig4}(c) and Fig.~\ref{fig4}(g). We find that it has no
gap and is linear in the vicinity of the minimum $\Gamma$ point. In
contrast, for $\lambda>1$, the system enters the $Z$-AFM phase. Its
excitation spectrum is shown in Fig.~\ref{fig4}(d) and
Fig.~\ref{fig4}(h). Now, energy gaps open up at both momenta
$(\pm\pi, 0)$ and $(0, \pm\pi)$.

\subsection{$3\times 3$ SkX Phase}

Finally, we study the $3\times3$ skyrmion crystal phase, which is
located in the phase diagram between the $4\times 1$ spiral spin
phase and the $2\times 2$ VX phase. As mentioned at the end section
III, The spin configuration of this phase consists of nine
sublattices, which are characterized by an up-spin at the center and
others lying on the $XY$-plane. By the spin-wave theory, we
calculate its spin excitation spectrum. Our results are shown in
Fig.~\ref{fig4}(b) and Fig.~\ref{fig4}(f). Obviously, the spectrum
has also an energy gap. Interestingly, we find that the spin
fluctuations dramatically destabilize this classical ground state
and make the skyrmion crystal phase unstable in part of the
classical phase diagram determined by the Monte Carlo simulations
\cite{Cole}. It may imply the appearance of a novel state in the
concerned region. However, in order to determine the properties of
this state, one must carry on more sophisticated analysis.

\section{CONCLUSIONS}

In summary, we study the exotic magnetic phases of two-component
Bose gases with spin-orbit coupling on a deep square optical lattice
by using spin-wave theory. In particular, we systematically
investiagte the spin-wave excitation spectra for the
$XY$-ferromagnet,  spiral spin, spin vortex, and skyrmion crystal
phases. We find that, while the excitation spectrum of the vortex
crystal phase has no gap and is linear in the vicinity of  the
center of Brillouin zone, the spectra of the commensurate spiral
spin phase and the skyrmion crystal phase possess energy gaps.
Experimentally, these spin textures can be easily observed through
spin-resolved time-of-flight measurements \cite{Lin2}. We discuss
also the stability of the novel skyrmion crystal phase and show that
the possible existing regions for the skyrmion crystal phase are
greatly shrunk by the spin-wave excitations. It suggests the
emergence of a new state in these regions. Finally, we would like to
emphasize that similar spin textures, which are discussed in the
present work, can be also induced by the DM type of interactions in
some solid state materials. Therefore, we believe that the current
investigation should provide with further insights into the study of
these materials.

\begin{acknowledgments}

We would like to thank Dr. X. F. Zhang for many helpful discussions.
This work is supported by NCET, NSFC under grants Nos. 11074175,
10934010, 11374017, NSFB under grants No.1092009, NKBRSFC under
grants Nos. 2011CB921502, 2012CB821305, and NSFC-RGC under grants
No. 11061160490.
\end{acknowledgments}

\end{document}